\documentclass[twocolumn]{aastex631}

\received{Received date}
\revised{Revised date}
\accepted{Accepted date}

\submitjournal{ApJ}

\begin{document}

\title{Investigating the spectral nature  of gigahertz-peaked spectra pulsar candidates}

\correspondingauthor{Karolina Ro{\.z}ko}
\email{krozko@uz.zgora.pl}

\author[0000-0002-1756-9629]{K. Ro{\.z}ko}
\affiliation{Janusz Gil Institute of Astronomy\\
University of Zielona G\'ora \\
ul. Prof. Z. Szafrana 2, \\
65-516 Zielona G\'ora, Poland}

\author[0000-0001-9577-708X]{J. Kijak}
\affiliation{Janusz Gil Institute of Astronomy\\
University of Zielona G\'ora \\
ul. Prof. Z. Szafrana 2, \\
65-516 Zielona G\'ora, Poland}

\author[0000-0003-1824-4487]{R. Basu}
\affiliation{Janusz Gil Institute of Astronomy\\
University of Zielona G\'ora \\
ul. Prof. Z. Szafrana 2, \\
65-516 Zielona G\'ora, Poland}

\author[0000-0001-9212-3574]{P. Dabhade}
\affiliation{Astrophysics Division, National Centre for Nuclear Research\\ 
Pasteura 7, 02-093 Warsaw, Poland}

\author[0000-0003-0513-9442]{W. Lewandowski}
\affiliation{Janusz Gil Institute of Astronomy\\
University of Zielona G\'ora \\
ul. Prof. Z. Szafrana 2, \\
65-516 Zielona G\'ora, Poland}

\begin{abstract}
We report the measurements of  low radio frequency spectra of fourteen gigahertz-peaked spectra (GPS) pulsar candidates, between 300~MHz and 700~MHz, using the upgraded Giant Meterwave Radio Telescope. Combining newly collected measurements with archival results the spectral nature of each pulsar was examined using four different physical models: simple power law, broken power law, low-frequency turn-over power law and free-free thermal absorption. Based on this analysis, we confirm the GPS nature of five pulsars, three of them being new detections. In addition, one pulsar can be classified as having a broken power law spectrum, and we found the typical power law spectra in four other cases. In the remaining four pulsars the spectra showed tendencies of low frequency turn-over that require further investigations at lower frequency ranges.  These results demonstrate the effectiveness of wideband measurements at low frequencies, below 1 GHz,  in characterizing  the spectral nature in pulsars. Our results also underline the need for more systematic theoretical studies to refine existing models and better interpret pulsar emission properties.

\end{abstract}


\section{Introduction}\label{sec:intro}
There is increasing observational evidence that the coherent radio emission from pulsars are excited in relativistic electron-positron pair plasma due to curvature radiation from charge bunches \citep{2000Melikidze, 2004Gil, 2009Mitra, 2023Mitra, 2024Johnston}. The observed power law spectra seen in a majority of pulsars arise due to incoherent averaging of the spectra from a large number of charge bunches, which further gets modified due to propagation of the radiation in the birefringent pair plasma medium with strong magnetic field \citep{2022Basu_a, 2024Mitra}. However, there are several pulsars whose spectra show deviation from the inverse power law dependence, which is apparently not associated with the intrinsic emission mechanism. In the early years after the discovery of pulsars, \citet{1973Sieber} proposed four models to characterize the shape of pulsar spectra: a simple power law, synchrotron self-absorption (considered as an analytical description rather than a physical interpretation), free-free thermal absorption in the interstellar medium (ISM), and a dual-power law model with two different spectral indices on either side of a break frequency. Early observations of turnover in the spectral shape at low frequencies (i.e. with a peak frequency $\leq 150$~MHz) revealed the existence of a correlation between spectral indices below and above the peak frequency, namely that the maxima of the spectra of different pulsars were observed to be rather symmetrical in shape, which was interpreted as an argument that some internal mechanism and not external factors are responsible for the observed changes in the spectrum \citep{1978Bruk}. The hypothesis about the intrinsic nature of low-frequency turnovers was further strengthened by the fact that observations of 30-millisecond pulsars (MSPs) showed that their spectra, unlike those of normal pulsars at that time, even at low frequencies followed a simple power law function \citep{2001Kuzmin}. Recently, \citet{2024Sharan} indicate a low-frequency turnover in MSPs. On the other hand, \citet{1979Malov} investigated the possibility of turnovers due to absorption in the pulsar plasma and proposed that such turnovers are the result of absorption by thermal electrons in the pulsar emission zone. A formula for the optical depth ($\tau$) was derived in this work and was later adopted by \citet{1981Izvekova} to model the observed low-frequency spectral turnovers in pulsars.

The number of pulsars with sensitive flux density estimates over a wide radio frequency range have significantly increased, and this led to a new class of pulsars being identified by \citet{2007Kijak, 2011KijakA} that showed high-frequency turnovers around 1 GHz and was called the gigahertz-peak spectra (GPS) pulsars. Over the following years, detailed studies helped distinguish them as a separate group within the pulsar population \citep{2013Allen,2014Dembska, 2015Lewandowski, 2016Rajwade, 2017Jankowski, 2017Kijak,2021Kijak, 2018Basu, 2018Rozko, 2020Rozko, 2021Rozko}. An important insight into this phenomenon was obtained from observations of the pulsar B1259$-$63 \citep{2011KijakB, 2015Dembska, 2018Koralewska}, which is part of a binary system with the Be star LS 2883 with a highly elliptical orbit. The spectrum of PSR B1259$-$63 evolves according to the pulsar's orbital phase around the Be star: showing a simple power law spectrum far away from the companion that morphs into a broken spectrum and subsequently a spectral turnover is seen as the pulsar gradually hides behind the wind ejected by its companion. The only absorption model that can work in such an environment is the free-free thermal absorption by the free electrons in the stellar wind or in the circumstellar disc of the Be star.
The free-free thermal absorption, first proposed by \citet{1973Sieber} for the low-frequency turnover, has emerged as the most likely mechanism for GPS behavior, with possible absorbing medium being either the material around neutron stars or the high density features in the ISM along the line of sight \citep[see for example][]{2013Kijak}. Likely candidates for the absorbing medium include dense supernova filaments, the material in the bow-shocks of pulsar wind nebulae and H\,{\sc ii} regions, and the physical properties necessary for explaining the spectral shape, in each of these cases, have been extensively studied \citep{2015Lewandowski, 2016Basu, 2016Rajwade}. The parameters of the thermal absorption model used to obtain the observed turnover in the spectra are usually insufficient to constrain the nature of the absorbing medium. However, under the assumption that the free electrons in the absorbing medium contribute to half the dispersion measure (DM), it was possible to eliminate candidate absorbing sources requiring non-physical parameters for the spectral turnover \citep{2017Kijak, 2021Kijak}.

Over the last ten years, many observational projects have been carried out to better understand the spectra of pulsars. Some of the prominent studies include: the LOFAR (LOw Frequency ARray) pulsar census \citet{2016Bilous,2020Bilous, 2016Kondratiev, 2020Bondonneau}, the pulsar flux density measurements from The LOFAR Tied-Array All-Sky Survey \citep[LOTAAS;][]{2019Sanidas}, the results of the Giant Metrewave Radio Telescope (GMRT) Sky Survey \citep[TGSS ADR and MSPES surveys][]{2016Frail,2021Basu}, multi-subbands pulsars flux density measurements obtained from the Parkes 64 m telescope \citep{2016Han, 2017Jankowski, 2018Johnston}, The Green Bank North Celestial Cap (GBNCC) pulsar census \citep{2020McEwen}, the Murchison Widefield Array pulsars flux densities \citep{2017Murphy}, The Southern-sky MWA Rapid Two-metre (SMART) pulsar census data \citep{2023Bhat}, the initial result from the Five-hundred-meter Aperture Spherical radio Telescope (FAST) Galactic Plane Pulsar Snapshot (GPPS) survey \citep{2021Han}, pulsar flux density measurements from the MeerKAT Thousand Pulsar Array programme \citep{2021Posselt} and MeerKAT Pulsar Timing Array \citep[MPTA;][]{2022Spiewak, 2023Gitika}. Recent studies of the behavior of pulsar spectra over a large population covering a large frequency range using narrow-band observation, have been carried out by \citet{2017Jankowski}. They analyzed 441 pulsars and categorized their spectral nature under five distinct classes available in the literature, namely: simple power law, broken power law, log parabolic spectrum, power law with high-frequency cut-off and power law with low-frequency turnover in a form proposed by \citet{1981Izvekova}. The same approach was later adapted by \citet{2022Swainston} in their spectral fitting software\footnote{\textit{pulsar\_spectra}: https://all-pulsar-spectra.readthedocs.io/en/latest/spectral\_index\_summary.html}  to carry out statistical analysis of 618 pulsars, including 59 MSPs.  

In this publication, we present the results of wideband observations of 15 GPS pulsar candidates in the frequency range 317-725~MHz, using the upgraded Giant Meterwave Radio Telescope (uGMRT). Our main goal is to investigate the GPS nature of these sources by comparing four spectral models, namely simple power law, broken power law, free-free thermal absorption and low-frequency turnover. The observational details and description of data analysis are reported in Section~\ref{sec:obs_and_an}. The measured flux densities and
estimates of the spectral modeling of the spectra of each pulsar are presented in Section~\ref{sec:res}, while in Section~\ref{sec:disc} we briefly discuss the implications of the results obtained in this work.    

\section{Observations and data analysis} \label{sec:obs_and_an}

\subsection{Observations} \label{subsec:obs}

We have carried out observations using the upgraded Giant Metrewave Radio Telescope (uGMRT) located near Pune, India \citep{1990Swarup,1991Swarup}. It is an aperture synthesis array consisting of 30 fully steerable antennas, each of 45 m diameter, with a maximum baseline of 25 km. For many years, GMRT was a narrow-band instrument that allowed observations at five different frequency ranges, but its new upgraded receiver system provides near-continuous coverage at four wide frequency bands: 120–250~MHz (band-2), 250–500~MHz (band-3), 550–850~MHz (band-4), and 1050–1450~MHz \citep[band-5;][]{2017Gupta,2019Raut}. 

The observations were conducted over three different cycles, three pulsars observed between May and June 2018 (project code : 34\_027), three pulsars observed between January and February 2021 (project code : 39\_065) and an additional nine pulsars observed between June and July 2023 (project code: 44\_060). In each case, the observing scheme was very similar. We simultaneously observed each pulsar with both available receiver systems: the GSB (the legacy GMRT Software Backend) at central frequencies 325 MHz and 610 MHz, and the GWB (the GMRT Wideband Backend) at band-3 (250–500~MHz) and band-4 (550–850~MHz). A total of 2048 spectral channels over the entire frequency band were recorded during the wideband observations and subsequently divided into $\sim 33$~MHz subbands. For GSB receiver we recorded over 33 MHz band-width, divided into 256 channels, at central frequencies of 325~MHz and 610~MHz. At the start and the end of each observational session, the flux calibrators 3C~286 and 3C~48 were observed to calibrate the flux density scale. Additionally, a suitable phase calibrator 1822$–$096 or 1714$–$252 was observed at regular intervals to correct for temporal variations and fluctuations in the frequency band. All pulsars were observed for $\sim 1$~hour each over two observational sessions, separated by a few weeks, to take into account the possible influence of interstellar scintillations\footnote{The diffraction scintillations should average out within about an hour of observation \citep{2012Lorimer}.}. 

\subsection{Data analysis} \label{subsec:data}

We employed standard image analysis schemes described in \citet{2021Rozko}, where the removal of bad data, the calibration of antenna response, and the image analysis were carried out using the Astronomical Image Processing System \citep[\textit{AIPS};][]{1990Greisen,2003Greisen}. The flux scales of the calibrators 3C~286 and 3C~48 were set using the \citet{2013Perley} estimates. A number of automated imaging pipelines have become available in recent years for faster processing of imaging data \citep[e.g,]{2014Intema,2016vanWeeren,2021Kale}. In this work, we have also utilized the automated imaging scheme, Source Peeling and Atmospheric Modeling \citep[\textit{SPAM}]{2014Intema,2017Intema}, that has the potential to simplify and greatly expedite the image analysis in future works. We intend to compare the results of the flux density measurement of the \textit{SPAM} pipeline with the standard analysis scheme to ensure consistency as well as have additional verification of our previous flux density estimates. Generally, the \textit{SPAM} pipeline is divided into two stages: the `pre-calibrate targets' step which includes, among others, the removal of channels affected by RFI, determination of flux scale and production of cross-calibration tables, and the `process target' step, that uses a self-calibration procedure to produce an image with corrections for ionospheric effects. More details about the ionospheric calibration can be found in \citet{2009Intema}. 
The details of the analysis using the \textit{SPAM} pipeline can be summarized as follows:
\begin{itemize}
    \item Step 1: Analysis of GSB data with a cutoff in the uv-range below  2 k$\lambda$ at both 325 MHz and 610 MHz frequencies, to ensure that the contributions of the extended sources are minimized and the point-like sources are clearly identified.  
    \item Step 2: Extraction of the reference sky model using the PyBDSF subroutine from the final image obtained in step 1. 
    \item Step 3: Band 3 of the GWB data was divided into six sub-bands (centered at frequencies 317, 350, 385, 417, 450 and 482~MHz) and band-4 into four sub-bands (centered at frequencies 578, 625, 675 and 725~MHz). Each sub-band was subsequently analyzed using the \textit{SPAM} pipeline in a manner identical to the narrow-band GSB data, but in the \textit{SPAM} main pipeline run was the reference sky model obtained for GSB data. For consistency in this step, we also applied a $uv$-range cutoff below 2 k$\lambda$, which allowed us to remove the interference pattern at the larger angular scales from the image.
    \end{itemize}

We have encountered flux scaling issues in our previous studies with GMRT \citep{2018Rozko,2021Rozko}. This is related to the phase calibrator 1822-096 whose flux density levels at certain times are not properly scaled by the Flux calibrator, likely due to their distant positions in the sky. In one of our earlier observations using the GSB the flux scale was set using 3C~48 at the start of the observation and 3C~286 at the end, which resulted in a factor of 3-4 variation in the flux levels of 1822-096. We resolved this issue by using the measurements of 1822-096 that was closest to the expected values at 325 MHz and 610 MHz for scaling the pulsar flux levels. In the observations reported in \citet{2021Rozko} using the wideband measurements of uGMRT, we once again encountered issues with the flux scaling of 1822-096 during one observing session on 30 May, 2018 at band-3 (300-500 MHz), which was lower by a factor of 2-3 compared to a second observing session 2 May, 2018. On this occasion the scaling issue was seen for both the Flux calibrators 3C~48 and 3C~286, and for six separate sub-bands across band-3. We carried out additional checks by comparing the flux levels of strong sources ($\sim$10) in the field of view and found similar discrepancy in the flux levels between the two days. We used the flux values from 2 May, 2018 for the final analysis.

In this present work using the standard imaging techniques it was found that the flux scale of the GSB measurements of 1822-096 were a factor of 3-4 times higher than the GWB observations in adjacent sub-bands, resulting in the estimated flux densities to be lower by the same factor. For example, PSR J1840$-$0809 had flux measurements of $3.67 \pm 0.48$~mJy at 325~MHz of the GSB system, but we obtained values of $11.6 \pm 1.5$~mJy and $14.3 \pm 3.7$~mJy at the 317~MHz and 344~MHz sub-bands of the GWB, respectively.

The \textit{SPAM} pipeline seems to have resolved this discrepancy and found consistent results between the GSB and the adjacent sub-bands of the GWB. The flux density of the pulsar J1840$-$0809 using the pipeline is $12.53 \pm 1.80$~mJy at 325~MHz band of GSB, while the adjacent GWB sub-band estimates are $10.9 \pm 1.2$~mJy at 317~MHz and $9.97 \pm 1.33$~mJy at 350~MHz. This clearly demonstrates that the SPAM pipeline gives similar results with the standard imaging technique, at least for the GWB observations. On the other hand it also suggests that the GSB measurements using standard imaging technique can be underestimated by factors of 3 to 4 from time to time due to inadequate estimation and transfer of the flux scaling from the calibrator to the target source. It should be emphasized that the \textit{SPAM} pipeline uses a different low frequency flux model, derived from \citet{2012Scaife}, and does not employ a direct scaling of the flux density levels of the source using the Flux calibrator.

In order to obtain consistent results we have reanalyzed the three pulsars observed in project 34\_027 \citep{2021Rozko} using the \textit{SPAM} pipeline, and reported the flux density measurements along with the newer observations of eleven pulsars in uGMRT observing projects 39\_065 and 44\_060. We also intend to use the pipeline as the primary measurement technique in our future studies.

\section{Results} \label{sec:res}
The fourteen pulsars studied in this work and \citet{2021Rozko} have been considered as GPS candidates based on unpublished narrow-band GMRT flux density measurements or on archival data. Pulsars J1741$-$3016, J1757$-$2223 and J1845$-$0743 had narrowband GMRT flux density measurements at frequencies 325~MHz, 610~MHz and 1280~MHz, which indicated the possible occurrence of turnover. These sources could not be associated with any potential absorbing medium in their vicinity and as a result, good spectral coverage was crucial to narrow down the physical parameters of possible absorbers. Preliminary analysis of their uGMRT wideband observations has already been published in \citet{2021Rozko}, but as mentioned in section~\ref{subsec:data}, we have improved our approach to solving the flux calibration problem and present the updated results here.

\begin{table*}[ht!]
	\centering
	\caption{Pulsars Flux Measurements at Band-3 from \textit{SPAM} pipeline for projects: 34\_027, 39\_065 and 44\_060; * means from only one observational session, $\leq $ means upper limit ($3 \sigma$), --- means that we don't have map from \textit{SPAM} pipeline. }
	\label{tab:pulsarsfluxes}
	\begin{tabular}{cccccccc} 
		\hline
	    Pulsar name & Project ID & S$_{317}$ & S$_{350}$ & S$_{385}$ & S$_{417}$ & S$_{450}$ & S$_{482}$ \\
	     & & mJy & mJy & mJy & mJy & mJy & mJy  \\
		\hline
	J1741-3016 & 34\_027 & $6.16 \pm 2.33$ & $8.15 \pm 2.64$* & $3.47 \pm      1.08$ & $4.86 \pm 1.13$ & $4.52 \pm 0.81$ & $4.14 \pm 0.72$ \\
	J1757$-$2223 & 34\_027 & $\leq 1.29$ & $\leq 1.11$ & $\leq 1.08$ & $\leq 0.91$ & $1.16 \pm 0.32$ & $\leq 1.91$ \\
  	J1845$-$0743 & 34\_027 & $5.84 \pm 0.74$ & $5.46 \pm 0.57$ & $5.66 \pm     0.48$ & $6.52 \pm 0.40$* & $5.37 \pm 0.25$ & $5.90 \pm 0.44$ \\
         & & & & & & \\
         J1827$-$0750 & 39\_065 & $9.18 \pm 0.63$ & $9.45 \pm 0.64$ & $9.23 \pm 0.59$ & $8.25 \pm 0.39$ & $8.46 \pm 0.49$ & $7.81 \pm 0.69$ \\
         J1840$-$0809 & 39\_065 &  $10.15 \pm 0.69$ & $9.51 \pm 0.67$ & $9.64 \pm 0.67$ & $8.88 \pm 0.42$ & $7.76 \pm 0.34$ & $7.17 \pm 1.19$*\\  
        J1841$-$0345 & 39\_065 & $\leq 1.73$ & $\leq 1.53$ & $\leq 1.42$ & $\leq 0.85$ & $\leq 1.09$ & --- \\
        & & & & & & \\   
        J1727$-$2739 & 44\_060 &  $5.8 \pm 1.7$ & $5.02 \pm 0.82$ & $5.1 \pm 1.4$ & $4.2 \pm 2.6$ & $4.16 \pm 0.36$ & $3.95 \pm 0.46$\\
        J1759$-$3107 & 44\_060 & $8.53 \pm 0.62$ & $8.76 \pm 0.53$ & $7.78 \pm 0.55$ & $6.99 \pm 0.85$ & $5.90 \pm 0.30$ & $5.85 \pm 0.68$ \\
        J1808$-$2057 & 44\_060 & $72.3 \pm 2.2$ & --- & --- & $ 45.2 \pm 1.1$ & $36.7 \pm 1.0$ & $27.1 \pm 4.3$ \\ 
        J1812$-$1733 & 44\_060 & $26.6 \pm 1.5$ & $27.0 \pm 1.5$ & $26.6 \pm 1.5$ & $22.38 \pm 0.68$ & $21.6 \pm 2.0$ & $22.53 \pm 0.95$ \\
        J1812$-$2102 & 44\_060 & $5.87 \pm 1.15$ & $4.86 \pm 0.83$ & $4.10 \pm 0.77$ & $4.66 \pm 0.42$ & $4.18 \pm 0.61$ & $3.69 \pm 0.50$ \\ 
        J1828$-$1101 & 44\_060 & $4.1 \pm 1.1$ & $4.6 \pm 1.1$ & $7.3 \pm 1.0$ & $6.23 \pm 0.60$ & $4.75 \pm 0.46$ & $5.31 \pm 0.80$ \\
        J1834$-$0731 & 44\_060 & $13.9 \pm 1.5$ & $12.0 \pm 1.6$ & $20.41 \pm 2.4$ & $6.15 \pm 0.85 $ & $9.7 \pm 1.4$ & $7.6 \pm 1.2$ \\
        J1845$-$0434 & 44\_060 &  $1.5 \pm 1.2$ & $3.3 \pm 1.2$ & $2.6 \pm 2.2$ & $2.96 \pm 0.62$ & $2.89 \pm 0.46$ & $2.39 \pm 0.59$ \\
		\hline
	\end{tabular}
\end{table*}

\begin{table*}[ht!]
	\centering
	\caption{Pulsars Flux Measurements at Band-4 from \textit{SPAM} pipeline for projects: 34\_027, 39\_065 and 44\_060; * means from only one observational session, symbol - means that we don't have a map from \textit{SPAM} pipeline. }
	\label{tab:pulsarsfluxes2}
	\begin{tabular}{ccccc} 
		\hline
	    Pulsar name & S$_{578}$ & S$_{625}$ & S$_{675}$ & S$_{725}$\\
	      & mJy & mJy & mJy & mJy \\
		\hline
		J1741$-$3016 & $7.66 \pm 0.86$ & $7.01 \pm 0.44$ & $6.43 \pm 0.79$ & $4.67 \pm 0.66$ \\
  J1757$-$2223 & $2.13 \pm 0.43$ & $2.62 \pm 0.74$* & $1.99 \pm 0.21$ & $1.74 \pm 0.22$  \\
    J1845$-$0743 & $6.37 \pm 0.41$ & $6.13 \pm 0.39$ & $6.05 \pm 0.33$ & $6.19 \pm 0.43$  \\
     & & & & \\
    J1827$-$0750 & $8.79 \pm 0.15$ & $8.45 \pm 0.12$ & $7.95 \pm 0.11$ & $7.64 \pm 0.12$ \\
    J1840$-$0809 & $6.97 \pm 0.19$ & $6.34 \pm 0.13$ & $5.77 \pm 0.13$ & $4.98 \pm 0.15$ \\
    J1841$-$0345 & $1.69 \pm 0.36$* & $0.96 \pm 0.22$* & $1.08 \pm 0.18$ & $1.57 \pm 0.22$ \\
         & & & & \\
    J1727$-$2739 & $2.50 \pm 0.15$ & $2.64 \pm 0.12$ & $2.41 \pm 0.64$ & $2.35 \pm 0.15$ \\
    J1759$-$3107 & $3.80 \pm 0.55$ & $3.2 \pm 2.6$ & $2.84 \pm 0.90$ & $2.46 \pm 0.99$ \\ 
    J1808$-$2057 & $22.6 \pm 3.1$ & $19.4 \pm 1.8$ & $15.4 \pm 2.5$ & $13.04 \pm 0.58$\\
    J1812$-$1733 & --- & --- & $12.9 \pm 1.3$ & $11.30 \pm 0.46$ \\
    J1812$-$2102 & $3.77 \pm 0.48$ & $3.31 \pm 0.62$ & $2.45 \pm 0.77$ & $2.49 \pm 0.48$ \\
    J1828$-$1101 & $6.28 \pm 0.37$ & $7.1 \pm 1.1$ & $4.9 \pm 1.2$ & $5.13 \pm 0.29$ \\
    J1834$-$0731 & $5.2 \pm 1.0$ & $7.6 \pm 1.5$ & $3.02 \pm 0.95$ & --- \\
    J1845$-$0434 & $2.8 \pm 1.0$ & $2.80 \pm 0.37$ & $2.97 \pm 0.64$ & $2.48 \pm 0.48$ \\
		\hline
	\end{tabular}
\end{table*}

Pulsars J1827$-$0750, J1840$-$0809 and J1841$-$0345 also had narrow-band GMRT flux density measurements at 325~MHz, 610~MHz and 1280~MHz that indicated possible turnover in their spectra. Additionally, for PSR J1841$-$0345 due to severe scattering, its flux density can only be securely measured from interferometric observations.

The remaining eight pulsars did not have any previous flux density measurements from GMRT but were candidates for GPS based on archival flux density values between 300 MHz and 800 MHz \citep[for example][]{1995Lorimer,2004Hobbs, 2005Champion, 2021Johnston}. Another argument was their large dispersion measure (DM) in the range $128-607$ pc/cm$^3$ \citep[see the ATNF Pulsar Catalogue,\footnote{http://www.atnf.csiro.au/people/pulsar/psrcat/}][]{2005Manchester}), which indicates the existence of potential areas with increased electron density in the line of sight. The spectral coverage in the archival data were relatively sparse, so the main goal of wideband observations was to understand the detailed spectral nature of these pulsars and assess whether they actually exhibit GPS behavior.

\subsection{Flux density measurements} \label{subsec:flux}
We intend to verify if the observed pulsars exhibit a GPS nature, hence it is crucial to obtain flux density measurements around the expected peak frequency. In Table~\ref{tab:pulsarsfluxes} we report the flux density values of the pulsars for six sub-bands in Band-3 and in Table~\ref{tab:pulsarsfluxes2} for four sub-bands in Band-4. The flux density of every pulsar over the two different observing sessions were consistent within measurement errors, and the average values are reported in the tables. We were able to measure the flux density in almost all pulsars, even if limited to a few sub-bands, with the exception of PSR J1916$+$0748, which was below the detection limit across the entire frequency range, and has been excluded from the subsequent analysis.

\begin{table*}[ht!]
	\centering
	\caption{The fitting parameters of the best spectral model obtained from modified \textit{pulsar\_spectra} software. The parameters: $A, \alpha, B, \beta, \nu_{\mathrm{pk}}, \alpha_1, \alpha_2$ are explained in the section~\ref{subsec:model}. In all models $\nu_0$ is the constant reference frequency of 10 GHz.}
	\label{tab:fittingpar}
	\begin{tabular}{cccccccc} 
		\hline
	    Pulsar name  & Model name & A & $\alpha$ & B & $\beta$ & $\nu_{\mathrm{pk}}$  & AIC \\
	      & & & & & & MHz &  \\
		\hline
   	J1741-3016 & FFA & $0.02 \pm 0.01$ & $-2.6 \pm 0.34$ & $0.0041 \pm 0.0009$ & - & $659 \pm 38$  & 55.3 \\
	 & LFT & $0.22 \pm 0.04$ & $-8.0 \pm 7.99$ & - & $0.438 \pm 0.005$ & $666 \pm 2$ & 46.0 \\
	J1757$-$2223 & FFA & $0.07 \pm 0.04$ & $-1.6 \pm 0.3 $ & $ 0.0031 \pm 0.0009$ & - & $727 \pm 45$ & 23.0 \\
	 & LFT &  $0.06 \pm 0.02$ & $-3.5 \pm 0.4$ & - & $0.62 \pm 0.04$ & $745 \pm 50$  & 26.6 \\ 
        J1828$-$1101 & FFA & $0.10 \pm 0.05$ & $-1.6 \pm 0.2$ & $0.0015 \pm 0.0004$ & - & $515 \pm 42$  & 46.1 \\ 
        & LFT & $0.14 \pm 0.06$ & $-4.3 \pm 0.5$ & - & $0.47 \pm 0.03$ & $509 \pm 4$ & 44.6 \\
        J1841$-$0345 & FFA & $0.29 \pm 0.02$ & $-0.95 \pm 0.06$ & $0.003 \pm 0.001$ & - & $918 \pm 137$  & 42.9\\
          & LFT & $1.7 \pm 0.3$ & $-1.88 \pm 0.14$ & - & $0.36 \pm 0.02$ & $610 \pm 100$  & 47.6 \\  
         J1845$-$0434 & FFA & $ 0.6 \pm 0.4$ & $-0.8 \pm 0.4$ & $0.0013 \pm 0.0008$ & - & $669 \pm 79$  & 99.3\\ 
          & LFT & $ 32.2 \cdot10^6 \pm 7.1 \cdot 10^6$ & $-8.0 \pm 7.8$ & - & $0.218 \pm 0.001$ & $760 \pm 20$  & 95.1\\      
         & & & & & & \\    
        J1812$-$1733 &  FFA & $0.18 \pm 0.04$ & $-1.7 \pm 0.1$ & $0.0006 \pm 0.0001$ & - & $323 \pm 18$ & 73.1 \\
          & LFT & $387 \cdot 10^6 \pm 6.15 \cdot 10^6$ & $-6.30 \pm 0.05$ & - & $0.1498 \pm 0.0006$ & $167 \pm 4$  & 59.0 \\
        J1812$-$2102 & FFA & $0.08 \pm 0.26$ & $-1.6 \pm 0.2$ & $0.0012 \pm 0.0004$ & - & $ \pm $   & 49.0\\
         &  LFT & $99.6 \cdot 10^9 \pm 27.7 \cdot 10^9$ & $-7.42 \pm 0.08$ & - & $0.1518 \pm 0.0009$ & $341 \pm 1$ & 31.2\\
         J1827$-$0750 & FFA & $0.07 \pm 0.01$ & $-1.89 \pm 0.09$ & $0.0015 \pm 0.0002$ & - & $475 \pm 22$  & 61.1 \\
         & LFT & $2.3 \pm 0.2$ & $-8.0 \pm 7.3$ & - & $0.2647 \pm 0.0009$ & $454 \pm 5$  & 45.2 \\
        J1840$-$0809 & FFA  & $0.24 \pm 0.02$ & $-1.22 \pm 0.05$ & $0.00036 \pm 0.00009$ & - & $312 \pm 34$ & 43.5\\
         & LFT & $1280 \pm 120$ & $-4.57 \pm 0.03$ & - & $0.1480 \pm 0.0005$ & $146 \pm 3$  & 41.2\\
            & & & & & & \\                 
        J1727$-$2739 & SPL & $0.2 \pm 0.1$ & $-0.9 \pm 0.2$ & - & - & -  & 25.2\\
        J1759$-$3107 & SPL & $0.14 \pm 0.09$ & $-1.22 \pm 0.03$& - & - & -  & 41.3\\
        J1808$-$2057 & SPL & $0.05 \pm 0.07$ & $-2.15 \pm 0.05$ & - & - & -  & 61.2 \\ 
        J1834$-$0731 & SPL & $0.040 \pm 0.003$ & $-1.68 \pm 0.05$ & - & - & -  & 34.7 \\
        		\hline
        & & & & & & \\ 
%\         J1845$-$0743 & $2.0 \pm 0.3$ & $-6.1 \pm 0.1$ & - & $0.262 \pm 0.001$ & $560 \pm 10$ & LFT & 45.2\\
%\         J1845$-$0743 & $0.35 \pm 0.08$ & $-1.2 \pm 0.1$ & $0.0012 \pm 0.0002$ & - & $531 \pm 24$ & FFA & 71.0\\
        Pulsar name & Model name & A & $\alpha_1$ & $\alpha_2$ & - & $\nu_{\mathrm{br}}$  & AIC\\
        		\hline
          	J1845$-$0743 & BPL &  $9.0 \pm 4.0$ & $0.14 \pm 0.15$ & $-1.8 \pm 0.1$& -&  $974 \pm 5$ & 27.8 \\
		\hline
	\end{tabular}
\end{table*}

\subsection{Spectral modeling} \label{subsec:model}

\begin{figure*}[ht!]
\centering
\includegraphics[width=2.1\columnwidth]{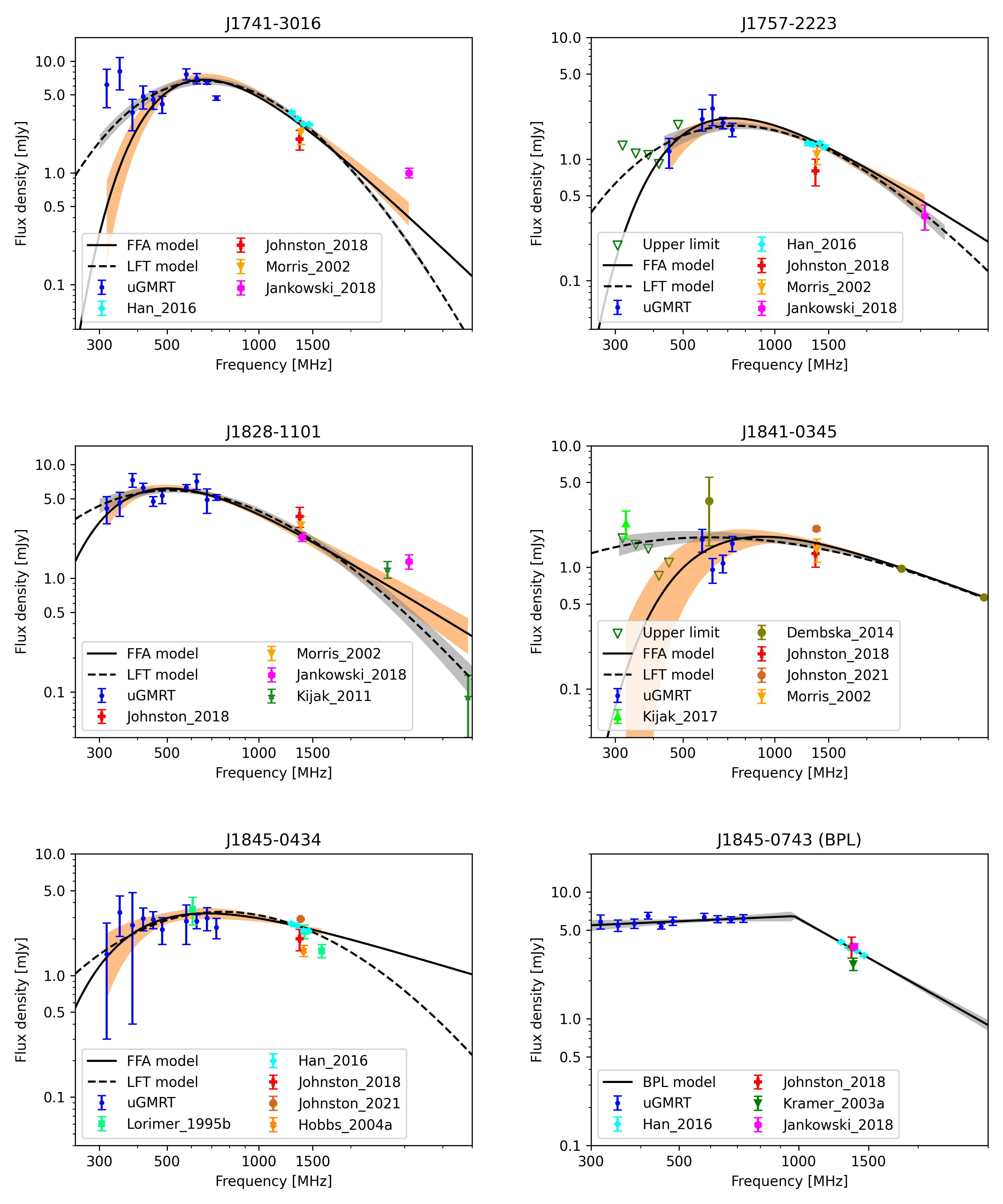}
\caption{The figure shows pulsars whose spectra are classified as gigahertz-peaked spectra (GPS). The solid line corresponds to the fitted free-free thermal absorption model and the orange color corresponds to $1\sigma$ contour. The dashed line shows the fit of the low-frequency turnover model and the gray color corresponds to $1\sigma$ contour. The spectrum of PSR J1845$-$0743 is classified as broken power law.  
\label{fig:spec1}}
\end{figure*}

\begin{figure*}[ht!]
\centering
\includegraphics[width=2.1\columnwidth]{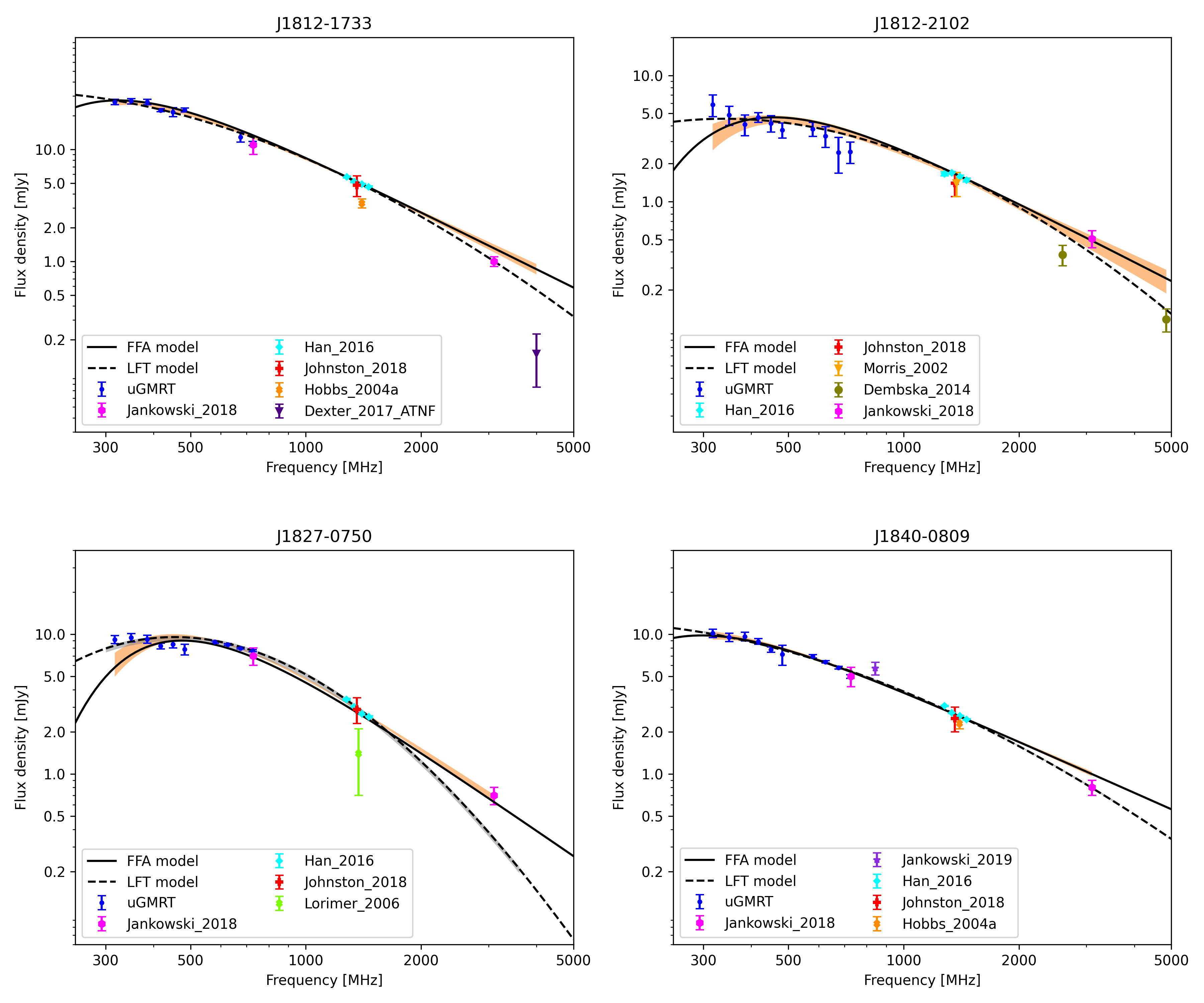}
\caption{The figure shows spectra of pulsars that are candidates for low-frequency turnover spectra (LFT).
\label{fig:spec2}}
\end{figure*}

\begin{figure*}[ht!]
\centering
\includegraphics[width=2.1\columnwidth]{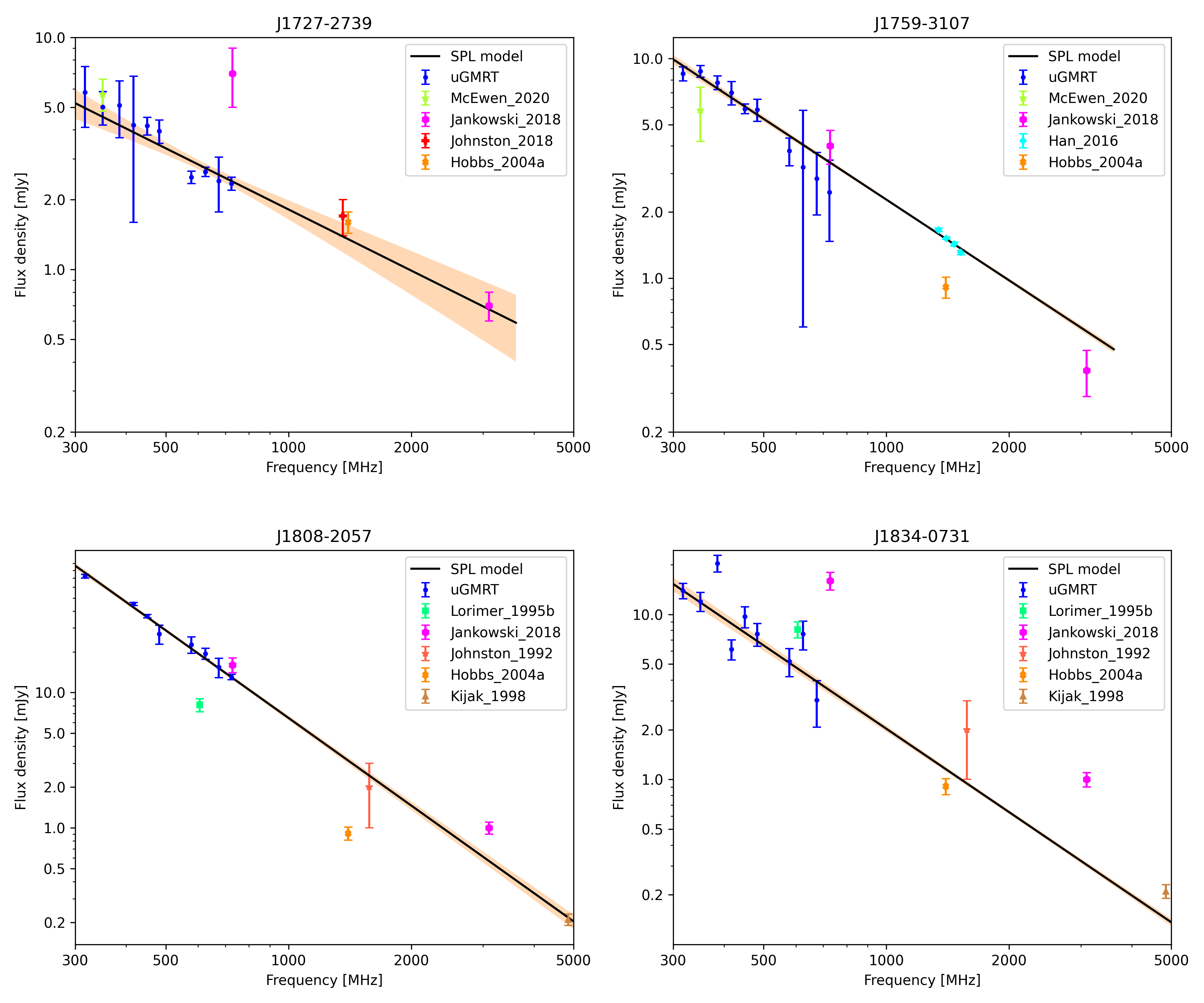}
\caption{The figure shows spectra classified as simple power law (SPL). 
\label{fig:spec3}}
\end{figure*}

We used the open-source, automated spectral fitting software, \textit{pulsar\_spectra} \citep[version 2.0.4]{2022Swainston}, to test four different functional forms as likely models for the pulsars in our sample, as described below:
\begin{itemize}
    \item Simple power law (SPL) function:
    \begin{equation}
    S_{\nu} = A \left( \frac{\nu}{\nu_0} \right)^{\alpha}
    \end{equation}
    where $A$ is constant, $\alpha$ is the pulsar spectral index and $\nu_0$ is reference frequency, fixed as 10 GHz for all fits.
\item Broken power law (BPL) function:
\begin{equation}
        S_{\nu} = A \left\{  \begin{array}{lc}
        \left(\frac{\nu}{\nu_0}\right)^{\alpha_1} & \mathrm{if}~  \nu \leq \nu_{\mathrm{br}} \\
        \left(\frac{\nu}{\nu_0}\right)^{\alpha_2} \left(\frac{\nu_0}{\nu_{\mathrm{br}}}\right)^{\alpha_1-\alpha_2} & \mathrm{otherwise} 
    \end{array} \right.,
\end{equation}
where $A$ is constant, $\nu_{\mathrm{br}}$ is the break frequency, and $\alpha_1$ and $\alpha_2$ are the spectral indices before and after $\nu_{\mathrm{br}}$, respectively.
\item Low-frequency turn-over (LFT) power law function: 
\begin{equation}
    S_{\nu} = A \left( \frac{\nu}{\nu_0} \right)^{\alpha} \exp{\left[ \frac{\alpha}{\beta}\left(\frac{\nu}{\nu_{\mathrm{pk}}}\right)^{-\beta} \right]}
\end{equation}
where $A$ is constant, $\alpha$ is a spectral index, $\nu_{\mathrm{pk}}$ is a turnover frequency and $0 < \beta < 2.1$. 
\item Free-free thermal absorption (FFA) model (the original software has been updated to include this model):
\begin{equation}
 S_{\nu} = A \left( \frac{\nu}{\nu_0} \right)^{\alpha} \exp{\left[-B \left(\frac{\nu}{\nu_0}\right)^{-2.1}\right]} 
\end{equation}
where $A$ is the intrinsic flux density at 10 GHz, $\alpha$ is the intrinsic spectral index and $B$ equals $0.08235 \times T_{\mathrm{e}}^{-1.35}~\mathrm{EM}$, EM is emission measure and $T_{\mathrm{e}}$ is temperature of the absorber.  
\end{itemize}
\textit{Pulsar\_spectra} software follows the fitting procedure described by \citet{2017Jankowski}, namely, it uses the Huber loss function to penalize outliers during the fitting procedure and uses the Akaike information criterion \citep[AIC; ][]{1974Akaike} to compare different spectral models. 

Table~\ref{tab:fittingpar} lists the values of the parameters obtained after fitting the relevant model to the measured spectra. The pulsars in our sample can be divided into the following three categories: 
\begin{enumerate}
    \item Pulsars whose spectra deviate significantly from a simple power law nature (see Figure~\ref{fig:spec1}). For all pulsars in this group, the peak frequency lies above 500~MHz, so we classify them as gigahertz-peaked spectra pulsars, with the exception of PSR J1845$-$0743, which clearly has a broken power law spectrum.
    \item Pulsars whose spectral behavior at low frequencies (i.e. below 300~MHz) is still unclear and therefore requires further study (see Figure~\ref{fig:spec2}).  
    \item Pulsars whose spectra follow simple power law function (see Figure~\ref{fig:spec3}). 
\end{enumerate}

\section{Discussion} \label{sec:disc} 
The absorption of radio emission in pulsars was initially proposed by \cite{1979Malov}, and subsequently adapted by \citet{1981Izvekova} to model the low-frequency turnover in the spectra. It was suggested that after the radio emission is generated due to curvature radiation in the pulsar magnetosphere with a typical power law spectral shape, they undergo absorption in the thermal electrons emitted from the heated surface of the neutron star. The pulsar flux density with free-free thermal absorption has the dependence :
\begin{equation}
    S = k \nu^\alpha \exp{\left(-\tau_{\nu}\right)}.
\end{equation}
Here, $k$ is the reference flux density, $\alpha$ is the intrinsic spectral index of emission, and $\tau_{\nu}$ is the optical depth of the thermal electron medium. The optical depth depends on the surface temperature, the strength of surface magnetic field, and the location of the emission region in the magnetosphere, specified by $\mathrm{b}R$, where $R$ is the radius of the neutron star and $\mathrm{b}$ is the emission height coefficient. The radio emission is expected to arise from a region well within the pulsar magnetosphere, i.e. $\mathrm{b}R << r_{\mathrm{L}}$, where $r_{\mathrm{L}} = P c/2\pi$ is the light cylinder radius. If one assumes that b is independent of frequency, then the model will be identical to the free-free thermal absorption proposed by \citet{1973Sieber} and later simplified by \citet{2013Wilson} and used in our work as the FFA model (see eq.4). On the other hand, when a frequency dependence of $\mathrm{b}$ is considered, namely $\mathrm{b} = b_0 \nu^{-\delta}$, where $\delta$ is the coefficient of frequency evolution of emission height, the expression has the form of power law with low frequency turnover, the LFT model in eq.(3). The expression for the optical depth ($\tau_{\nu}$) in this case has the form :
\begin{equation}
    \tau_{\nu} =  \left( \frac{\nu_{m}}{\nu} \right)^{2-5\delta} \frac{\alpha}{2-5\delta},
\end{equation}
where $\nu_{m}$ is peak frequency of spectral turnover.
\citet{2017Jankowski} uses a statistical coefficient $\beta$ in the LFT model which controls the curvature in the trailing part of the spectra (see discussion below). This term is identical to $2-5\delta$ in eq.(6). 
%where $\delta$ has a physical interpretation as the coefficient of frequency evolution of emission height.

Conversely, the synchrotron self-absorption, as described in \citet{1979Rybicki}, is a separate mechanism that has been applied to pulsar emission studies by \citet{1973Sieber}, based on the work of \citet{1961LRoux}, and has the analytical functional form:
\begin{equation}
    S = k \left( \frac{\nu}{\nu_0} \right)^{2.5} \left[ 1 - \exp{\left(-c \left(\frac{\nu}{\nu_0}\right)^{\alpha-2.5}\right)} \right],
\end{equation}
where $k, c$ and $\alpha$ can be obtained from spectral fits. This mechanism is different from the free-free thermal absorption model considered by \cite{1979Malov}, and has an exact dependence of $\nu^{2.5}$ for the inherent spectra in the optically thick regime. From the very inception, it was recognized that this model is not a good fit for pulsar emission, as it is primarily applicable to incoherent emission in small magnetic fields. While the pulsar emission mechanism is coherent in nature (see section 1) and emerges from the inner magnetosphere which is dominated by large magnetic fields.

The thermal absorption model of \cite{1979Malov} assumes that low-frequency turnover is caused by thermal electrons emitted from the surface. The pulsar plasma is generated due to pair production in an inner acceleration region above the polar caps, where it gets accelerated to ultra-relativistic energies by the large electric fields \citep{1971Sturrock, 1975Ruderman,2022Basu_b}. Hence, it is not feasible to produce the required thermal electrons in this system. In the work of \citet{2017Jankowski}, there is general confusion in terminology, where it has been suggested that the LFT model is motivated by synchrotron self-absorption. 
As mentioned above, the LFT model follows from the free-free thermal absorption which is quite distinct from synchrotron self-absorption, and should not be confused to have the same physical origin.

\begin{figure}[ht!]
\centering
\includegraphics[width=1.0\columnwidth]{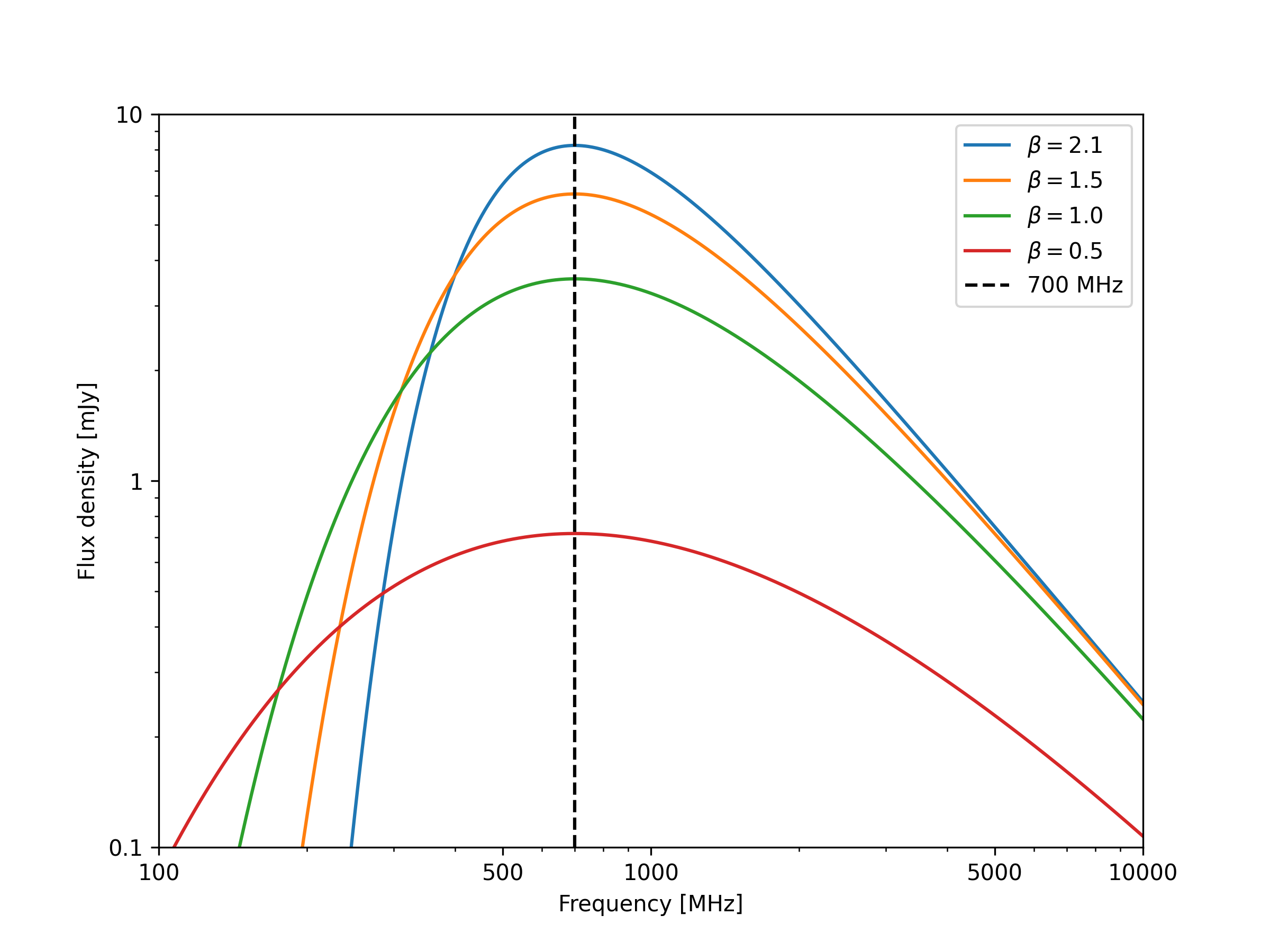}
\caption{The figure shows how the spectrum shape depends on different values of $\beta$ parameter in the low-frequency turnover model. The fixed values of other model parameters were: $A = 0.25$, $\alpha = -1.6$, $\nu_{\mathrm{pk}} = 700$~MHz, $\nu_0 = 10000$~MHz.  
\label{fig:beta}}
\end{figure}

In our previous works, we utilized the free-free thermal absorption model with the condition that pulsar radiation is absorbed somewhere along its path to the observer. Potential candidates for absorbers include: a dense filament of supernova remnant, a bow-shock pulsar wind nebula or H\,{\sc ii} region along the line of sight. To simplify the fitting process, we utilized the expression for optical depth as described by \citet{2013Wilson}:
\begin{equation}
\tau_{\nu} = 8.235 \times 10^{-2}  \left( \frac{T_{\mathrm{e}}}{K} \right)^{-1.35}  \left( \frac{\nu}{\mathrm{GHz}} \right)^{-2.1} \left( \frac{\mathrm{EM}}{\mathrm{pc~cm^{-6}}} \right) a(\nu,T),
\end{equation}
where $T_{\mathrm{e}}$ is the electron temperature, $\mathrm{EM}$ is the emission measure, and $a(\nu,T)$ is typically $\cong 1$. For computational efficiency, we assumed that at $\nu_0 = 10~\mathrm{GHz}$, absorption is negligible, reducing the number of free parameters in the spectral fits to three. This approach has the advantage of providing physically interpretable parameters that with additional physical constraints obtained from known properties of the candidate absorbers can be used to make reasonable inferences about the nature of absorbers \citep[see for example][]{2021Rozko, 2021Kijak}. However, it is worth noting the limitations, particularly the assumption of a uniform electron distribution within the absorbing medium. Incorporating a model for thermal absorption with an inhomogeneous electron distribution is necessary for greater accuracy but is beyond the scope of this work.

Regarding the LFT model discussed earlier (see eq.3), its underlying assumptions about the physical processes are no longer valid, but it retains practical advantages. For instance, the $\beta$ parameter in the model governs the smoothness of the spectral fit, allowing it to accommodate diverse spectral shapes. Figure~\ref{fig:beta} illustrates this adaptability, with other parameters held constant. For $\beta = 2.1$, the LFT model converges to the FFA model, as noted by \citet{2017Jankowski}, while smaller values of $\beta$ flatten the low-frequency portion of the spectrum, yielding more symmetrical shapes around the peak. This symmetry can mimic synchrotron self-absorption, which may be relevant in pulsars with pulsar wind nebulae (PWNe) where the compact nebula contributes significantly to the flux density \citep[see][]{2018Basu}.
   
Interestingly, spectral fits often yield unrealistically small intrinsic spectral indices ($\alpha \approx -6$ or $-8$) which are actually the fitted parameters boundaries set in fitting procedures for the LFT model. Assuming minimal absorption at $\nu_0 = 10~\mathrm{GHz}$, these values are nonphysical given that the mean spectral index for pulsars is around $-1.6$ \citep{2017Jankowski}.

To balance the strengths and weaknesses of both models, we considered both LFT and FFA models for our analysis. Our observations confirmed GPS behavior in five pulsars: J1741$-$3016, J1828$-$1101, J1845$-$0434, J1757$-$2223, and J1841$-$0345 (see Figure~\ref{fig:spec1}). Of these, J1741$-$3016 and J1757$-$2223 were previously classified as GPS pulsars \citep{2021Rozko}, while the remaining three were initially identified as simple power-law spectra \citep{2017Jankowski, 2021Johnston}. PSR J1845$-$0743 was previously classified as GPS pulsar \citep{2021Rozko}. After reanalyzing flux density measurements using \textit{SPAM} pipeline due to its relatively flat spectrum below the peak frequency, PSR J1845$-$0743 spectrum is classified as a broken power law.

For two pulsars, J1757$-$2223 and J1841$-$0345, the AIC criterion favored the FFA model, while the LFT model was preferred for the other three. However, the AIC differences were within 20\%. Notably, the spectral indices derived from the LFT model were smaller and, in two cases, reached the lower limit ($-8.0$), whereas the FFA model produced more realistic values. Table~\ref{tab:pulsarsproperties} summarizes key properties of the observed pulsars, including their dispersion measures (DM), which served as a selection criterion for GPS candidates. Despite our efforts, only four pulsars could be associated with potential absorbing media, highlighting the limitations of our sample. Ultimately, GPS features were confirmed in only about 30\% of the selected sources.

\begin{table*}[ht!]
	\centering
	\caption{The main physical parameters of observed pulsars. All values are taken from ATNF Catalog$^a$. The remarks column indicates possible associations of the neutron star with a supernova remnant (SNR) nebula, a pulsar wind nebula (PWN), an H\,{\sc ii} region, or an unidentified X-ray source from the HESS catalog. References: 1 - \citet{2018HESS}, 2 - \citet{2024Ocker}, 3 - \citet{2022Abdollahi},  4 - \citet{2013Stroh}}
	\label{tab:pulsarsproperties}
	\begin{tabular}{ccccccc} 
		\hline
	    Pulsar name & DM & Distance & Age & Spectral & Remarks & References\\
        & & & & classification & & \\
	      & cm$^{-3}$ pc & kpc & kyr &  & & \\
		\hline
        J1727$-$2739 & $146.0 \pm 0.3$ & $3.959$ & $19.1$ & SPL & & \\
	J1741$-$3016 & $382 \pm 6$ & $3.87$ & $3.34$ & GPS & HESS  J1741$-$302 ? & 1\\
        J1757$-$2223 & $239.3 \pm 0.4$ & $3.727$ & $3.75$ & GPS & &  \\
        J1759$-$3107 & $128.3 \pm 0.3$ & $3.1$ & $4.53$ & SPL &  & \\
        J1808$-$2057 & $606.8 \pm 0.9$ & $4.586$ & $0.852$ & SPL & &\\ 
        J1812$-$1733 & $509.8 \pm 0.1$ &  $4.492$ & $8.68$ & LFT & H\,{\sc ii}: S40 & 2\\
        J1812$-$2102 & $547.2 \pm 0.1$ & $14.443$ & $0.811$ & LFT & \\
        J1827$-$0750 & $375.45 \pm 0.07$ & $9.831$ & $2.77$& LFT & & \\
        J1828$-$1101 & $605.0 \pm 0.1$ & $4.767$ & $0.0772$ & GPS & GRS:4FGL\_J1828$-$1059 & 3\\
        J1834$-$0731 & $288.3 \pm 0.4$ & $4.071$ & $0.14$ & SPL & 4FGL J1834.7-0724c ? & 4\\
        J1840$-$0809 & $349.8 \pm 0.8$ & $5.745$ & $6.44$ & LFT & & \\
        J1841$-$0345 & $194.32 \pm 0.06$ & $3.776$ & $0.0559$ & GPS & & \\
        J1845$-$0434 & $230.8 \pm 0.2$ & $4.096$ & $0.681$ & GPS & & \\
        J1845$-$0743 & $280.93 \pm 0.02$ & $7.114$ & $4.52$ & BPL & &  \\
     & & & & \\
		\hline
	\end{tabular}
    {$^a$https://www.atnf.csiro.au/research/pulsar/psrcat/ \citep{2005Manchester}}
\end{table*}

\section{Conclusions} \label{sec:con}
Studying pulsar spectra across a wide range of frequencies presents significant challenges due to the diverse observational data collected using multiple radio telescopes and differing observation techniques. This complexity becomes even more pronounced when analyzing large samples of pulsars, where two primary approaches dominate:
\begin{enumerate}
    \item Fitting morphologically simple models, such as a single power law or power laws with one or two breaks \citep{2016Bilous,2020Bilous, 2017Murphy}.
    \item Exploring a broader set of models proposed in the literature, regardless of their physical motivations \citep{2017Jankowski, 2022Swainston}.
\end{enumerate}

In our study, we focused on pulsar spectra exhibiting turnover at relatively high frequencies, prioritizing physically meaningful models. We specifically tested two models for pulsars that clearly exhibits turnovers—the FFA model and the LFT model—on our sample of pulsars. While the AIC often favors the LFT model, the parameters derived from the FFA model align more closely with realistic physical interpretations of pulsar emission. Notably, the validity of the LFT model, as derived by \citet{1979Malov}, appears inconsistent with current theoretical understanding of pulsar emission. We believe that the theoretical derivation of this model warrants re-evaluation, though such an effort lies beyond the scope of this work.

We also reported flux density measurements from wideband observations of 15 candidate GPS  pulsars, conducted using uGMRT Band-3 and Band-4 receivers. Based on our analysis, we confirm the GPS nature in the spectra of five pulsars: J1741$-$3016, J1828$-$1101, J1845-0434, J1757$-$2223 and J1841$-$0345. Additionally, we classify the spectrum of PSR J1845$-$0743 as a broken power law.
 
These findings contribute to the broader understanding of pulsar spectra and highlight the importance of using physically motivated models in spectral analysis. Our results underline the need for more systematic theoretical studies to refine existing models and better interpret pulsar emission mechanisms. Future wideband observations and rigorous modeling will be key to unveiling the complex spectral behavior of pulsars, particularly those with GPS characteristics.

Radio Astronomy has entered a new era of wideband receiver systems with currently operational uGMRT, MeerKAT, Parkes UWL, FAST and upcoming SKA. It will be important to take into account all different kinds of spectral behavior of the expected sources while implementing the pulsar search algorithms in future studies. For example the Band 1 of SKA-Mid receiver is planned to have a frequency coverage between 350-1050~MHz \citep{2018Levin}, and should account for GPS and broken spectral nature otherwise a certain fraction of pulsars will be missed.

\section{Acknowledgments}
We thank the anonymous referee for constructive comments that helped to improve the paper. We thank Dipanjan Mitra for useful discussions about the physical origin of pulsar spectra. We thank the staff of the GMRT who have made these observations possible. The GMRT is run by the National Centre for Radio Astrophysics of the Tata Institute of Fundamental Research. 
This work was partially supported by a grant of the National Science Center, project no. 2020/37/B/ST9/02215.

\vspace{5mm}
\facilities{uGMRT}

\software{\textit{AIPS} software (version 31DEC18) \citep{1985Wells}, \textit{SPAM} pipeline \citep{2014Intema}, 
        \textit{pulsar\_spectra} (version 2.0.4) \citep{2022Swainston},
        }

\bibliography{2024_Rozko_et_al}{}
\bibliographystyle{aasjournal}

\end{document}